\documentclass[preprint,12pt,3p]{elsarticle}

\usepackage{lineno,hyperref}
\usepackage{comment}
\usepackage{longtable}
\usepackage{subfigure}
\usepackage{float}
\usepackage[table]{xcolor}
\usepackage{array,ragged2e}
\modulolinenumbers[5]

\usepackage{multibib}
\newcites{rv}{Review Refereces}

\newcolumntype{L}[1]{>{\raggedright\let\newline\\\arraybackslash\hspace{0pt}}m{#1}}
\newcolumntype{C}[1]{>{\centering\let\newline\\\arraybackslash\hspace{0pt}}m{#1}}
\newcolumntype{R}[1]{>{\raggedleft\let\newline\\\arraybackslash\hspace{0pt}}m{#1}}

\journal{None}

\bibliographystylerv{elsarticle-num}

\begin{document}

\begin{frontmatter}

\title{Software Engineers' Attitudes Towards Organizational Change - an Industrial Case Study}
\author[adrcth]{Per Lenberg\corref{cor1}}
\address[adrcth]{Div. of Software Engineering, Chalmers University, SE-412 96 Gothenburg, Sweden, +46734379815}
\ead{perle@chalmers.se}
\cortext[cor1]{I am corresponding author}

\author[adrpsy]{Lars G{\"o}ran Wallgren}
\address[adrpsy]{Dept. of Psychology, Gothenburg University, Gothenburg, Sweden}
\ead{larsgoran.wallgren@psy.gu.se}

\author[adrbth]{Robert Feldt}
\address[adrbth]{Dept. of Software Engineering, Blekinge Inst. of Technology, Karlskrona, Sweden}
\ead{robert.feldt@bth.se}

\begin{abstract}
In order to cope with a complex and changing environment, industries seek to find new and more efficient ways to conduct their business. According to previous research, many of these change efforts fail to achieve their intended aims. Researchers have therefore sought to identify factors that increase the likelihood of success and found that employees' attitude towards change is one of the most critical. 

The ability to manage change is especially important in software engineering organizations, where rapid changes in influential technologies and constantly evolving methodologies create a turbulent environment. Nevertheless, to the best of our knowledge, no studies exist that explore attitude towards change in a software engineering organization.

In this case study, we have used industry data to examine if the \emph{knowledge} about the intended change outcome, the understanding of the \emph{need for change}, and the feelings of \emph{participation} affect software engineers' \emph{openness to change} and \emph{readiness for change} respectively, two commonly used attitude constructs. The result of two separate multiple regression analysis showed that \emph{openness to change} is predicted by all three concepts, while \emph{readiness for change} is predicted by \emph{need for change} and \emph{participation}. In addition, our research also provides a hierarchy with respect to the three predictive constructs' degree of impact. It shows that \emph{knowledge} has nearly fifty percent more impact compared to \emph{participation} (only valid for \emph{openness to change}), and that \emph{participation}, in turn, has nearly fifty percent more impact compared to \emph{need for change}. 

Ultimately, our result can help managers in software engineering organizations to increase the likelihood of successfully implementing change initiatives that result in a changed organizational behavior. However, the first-order models we propose are to be recognized as early approximations that captures the most significant effects and should therefore, in future research, be extended to include additional software engineering unique factors.
\end{abstract}

\end{frontmatter}


\section{Introduction}
\label{sec_introduction}
In order to cope with a complex and changing environment, organizations seek to find new and more efficient ways to conduct their business~\cite{serour2006towards, platt2007software, serour2007radar, greenwood1996understanding}. The capacity to manage change has become a key determinant of competitive advantage and survival~\cite{d2010hypercompetition}, and, therefore, industries need to adopt and utilize new processes, technologies and innovations that may enable them to achieve their goals.

The ability to conduct and cope with organizational change is especially important in software engineering organizations, where rapid changes in influential technologies, difficulty of settling requirements up-front, the inherent flexibility of software, and constantly evolving and changing methodologies create a turbulent environment~\cite{highsmith2001agile, nerur2005challenges, lenberg2015human}. The employees in software engineering organizations are frequently exposed to organizational change, which is a considerable source of stress~\cite{ferrie1995health, woodward1999impact}. Thus, these changes need to be managed as smoothly as possible in order to maintain healthy stress levels and keep employees motivated.

Even if the importance of organizational change has been acknowledged, many of the change efforts fail to achieve their intended aims~\cite{beer2000breaking}. For example, a survey of over 3,000 executives reported that two thirds of the respondents indicated that their companies had failed to achieve an improvement after implementing organizational changes~\cite{meaney2008mckinsey}.

In response to the high failure rate, work and organizational researchers have sought to identify factors that increase the likelihood of successfully implementing organizational changes. The research has shown that one of the most critical factors is employees' attitude towards change~\cite{rafferty2013change, oreg2011change}. An organizational change cannot be considered successful without a change in the employees' behavior~\cite{kotter2002heart}, which, according to social psychology researchers~\cite{ajzen2001nature}, is controlled and predicted by attitudes.

To the best of our knowledge, no study has previously explored attitudes towards organizational change in software engineering organizations. Given the importance of attitudes, it is, both from a research and from a managerial point-of-view, of interest to understand what underlying factors or constructs affect them. Therefore, the primary purpose of this study was to create, verify and validate a first-order model\footnote{The order of approximation indicate how precise an approximation is. First-order approximation is the term use for a further educated guess at an answer.~\cite{wiki:ordersofapproximation}} that predicts software engineers' attitude towards organizational change.

In order to create the model, we used the \emph{traditional measurement development procedure}; a method developed for such purposes and proven within social science research~\cite{viswanathan2005measurement}. As suggested by these procedures, we first conducted a systematic literature review (SLR) in order to gain a better understanding of the domain and, also, to identify factors or concepts that have had a significant impact on organizational change in software engineering organizations. The results from the SLR are reported separately since no such review has previously been published.

By combining the software engineering domain knowledge, gained through the SLR, with existing organizational psychology change theories, we then compiled two first-order models and verified them using industrial data collected from a Swedish software development company currently undergoing an organizational change. The data were collected at a single occasion in the beginning of the change process and, therefore, we cannot, in this present study, make any statement regarding the outcome or success of that specific change.

In the next section, we give further background information regarding literature reviews, attitudes in general and also attitudes in relation to organizational change. Then, we present the method and result of the systematic literature review after which we present the method and result of the attitude case study. Finally, the result is concluded.

\section{Background and Related Research}
\label{sec_background}
In the following sections, we briefly describe previous research that we have deemed most relevant and that has affected our study. This includes work and organization psychology, behavioral software engineering, the attitude concept and, finally, attitude in relation to organizational change. In addition, we briefly describe the company in which we conducted the case study and, also, the organizational change that the company was undergoing.

\subsection{Work and Organizational Psychology}
Psychology is the study of the mind and behavior~\cite{apa2015definepsychology}. Naturally, organizational psychology\footnote{Also sometimes referred to as industrial and organizational psychology, occupational psychology, or work psychology.} is the application of psychology in the workplace, i.e. is concerned with `behavior in the workplace'~\cite{muchinsky1997psychology}.

Work and organizational psychology has been in existence for about the last century. The question of what is significant for an individual's well-being and job satisfaction has been one of the most important research areas in organizational psychology since the 1920s. In the 1920s the research concentrated on physical work conditions such as lighting, ventilation and noise level. In the beginning of the 1930s to the beginning of the 1940s, the interest in the social aspects of the work environment increased. During these years the “human relations” – movement began, with Elton Mayo (1946) as one of its main spokesmen. Today work and organizational psychology raises important questions about how to manage effectively in organizations, in particularly with the increasing number of knowledge workers whose commitment is critical to organizational success.

\subsection{Behavioral Software Engineering}
Lenberg, Feldt and Wallgren have defined the research area of Behavioral Software Engineering (BSE) as the study of cognitive, behavioral and social aspects of software engineering performed by individuals, groups or organizations~\cite{lenberg2014towards}. A BSE literature review~\cite{lenberg2015behavioral} indicated that the human aspect of software engineering is a growing area of research that has been recognized as important. However, the review also showed that there are knowledge gaps and that earlier research has been focused on a few concepts, which have been applied to a limited number of software engineering areas, and, also, that the BSE research, so far, rarely has been conducted in collaboration by researchers from both software engineering and social science.

\subsection{The Attitude Concept}
According to Ajzen~\cite{ajzen2001nature}, it is commonly accepted that attitude represents a summary evaluation of an object captured in dichotomous dimensions such as good-bad, harmful-beneficial, pleasant-unpleasant and likable-dislikeable. One frequently used attitude model is the expectancy-value model~\cite{fishbin1972consideration}, which basically states that the overall attitude towards an object is determined by the sum of the beliefs towards the same object. Each belief is weighted by the strengths of its constituting, individual beliefs. 

Furthermore, attitudes are considered an important area of research in social psychology since they predict behavior~\cite{crano2006attitudes}. The most prominent behaviour prediction model is the theory of planned behaviour (TPB) and, its somewhat less used predecessor, the theory of reasoned action (TRA). According to TPB, people act in accordance with their intentions and perceptions of control over the behaviour while intentions in turn are influenced by attitudes towards the behaviour, subjective norms and perception of behaviour control. Another acclaimed theory, also closely associated with attitude change, is the cognitive dissonance theory developed by Festinger~\cite{festinger1962theory} in the 1950’s. The theory states that people are motivated to reduce dissonance, which can be achieved through changing their attitudes or beliefs.

Regarding the formation and change of attitudes, Crano and Prislin~\cite{crano2006attitudes} state the there exists two types of process models, single or dual process models. The singe process model operates automatically while the dual process operates in a controlled fashion~\cite{gawronski2013dual}. The dual process models are the most influential and an example of such is the heuristic-systematic model (HSM), which describes two depths in the processing of attitude change: systematic and heuristic~\cite{johnson1989effects}. The level of process is, to a certain extent, determined by the level of motivation and/or cognitive ability, where systematic processing occurs when individuals are motivated and have a high enough cognition to process a message, and, consequently, heuristic processing occurs when the individuals have low motivation and/or low cognitive ability to process a message. Hence, when an individual is unmotivated or unable to process a message, they will use less cognitive intensive features to form the attitudes. The individual uses, what Crano and Preslin~\cite{crano2006attitudes} refer to as “peripheral cues” or heuristics (e.g., “Dad’s usually right”), which are more related to the source than the actual message content. In the latter case, the source will play a more important part in the attitude formation. However, it should also be noted that heuristic attitudes are less stable and less likely to influence behaviour, compared to those formed by systematic processing. Furthermore, when attitudes change Wilson et al.~\cite{wilson2000model} mean that the new attitude overrides but may not replace the old attitude. Instead, people can simultaneously hold two different attitudes toward a given object in the same context, one attitude implicit or habitual, the other explicit, where, yet again, motivation and cognitive ability are assumed to be required to retrieve the explicit attitude.

\subsection{Organizational Change and Attitudes}
Organization change is both the process in which an organization changes its structure, strategies, operational methods, technologies or organizational culture to affect change within the organization and the effects of these changes on the organization~\cite{juma2014organizational}. The study of change is a major topic in organizational sciences. According to Bouckenooghe~\cite{bouckenooghe2010positioning}, research into organizational change can be categorized into two main themes: exploring the antecedents and consequences of change, and exploring how organizational change develops, grows, and terminates over time. The first of the two themes primarily addresses two topics: way to convince employees to accept changes, and manage employees’ attitudes toward change.

Attitude researchers have used several different constructs in order to measure different facets of attitudes towards organizational change. In a literature review, Choi~\cite{choi2011employees} identified the following four constructs as the most common: readiness for change, openness to a change, commitment to change and cynicism about an organizational change. Choi concluded that these attitude constructs are susceptible to situational variables and may change over time as the individuals’ experiences change, and that they, therefore, are better conceptualized as states than as personality traits.

\subsection{Company and Organizational Change Description}
\label{sec_background_company}
The data in this current study was collected at a department within a large Swedish software development company that was planning to initiate an organizational change. The department, which developed safety critical software for the global market, wanted to change their development process and, more specifically, transfer from a project-driven development to a product-driven development.

Traditionally, the products that the department owned had, to a large extent, been developed by customer projects, i.e. projects responsible for developing, customizing and delivering the product to a single customer. A consequence of this was that the products' feature sets were determined by the needs of the current paying customers and, consequently, the products became tailored to the existing customers' needs, not to the needs of the potential market as a whole. In addition, the quality of the products was dependent on the projects' budget. If one project ran out of money, the product quality were affected negatively, which led to a gradual degrease in quality and made feature development increasingly more expensive.

The management wanted a transfer of power from the project managers to the product managers. Project managers are responsible for the successful delivery of a singe project. They align resources, manage issues and risks, and basically coordinate all of the elements necessary to complete the complete delivery. The project managers can undertake to build a product, to add new features to a product, or create new versions or extensions of a product. However, after the delivery, they move on to new projects that might involve different products. The product managers, on the other hand, are responsible for the overall and ongoing success of a product. Once the project to build the product is complete and the project manager has moved on, the product manager remains to manage the product through the entire life-cycle.

Hence, the purpose of the organizational change was to decrease single customer projects' influence on the product feature-set and the product quality. The idea was to group the developers in autonomous teams controlled by a product manager who had the product's long-term goals as guidance when making operational decisions, not the goals of a single project. The organizational change affected the organizational structure and also the software development processes.

\section{Organizational Change in Software Engineering - A Systematic Literature Review}
\label{sec_SLR}
The primary purpose of the systematic literature (SLR) review was, as stated in the introduction, to identify factors or concepts that have had a significant impact on the organizational change in software engineering organizations. In addition, the secondary purpose was to compile an overview of the current organizational change research related to software engineering. We wanted to identify gaps in the current research and research methods, identify trends and point to directions for future research.

\subsection{Method}
\label{sec_method}
The SLR was based on the guidelines described by Kitchenham~\cite{kitchenham2004procedures}. In order to reduce the possibility of researcher bias, a pre-defined review protocol, describing the review process, was produced. The process included the following stages: (1) analyzing the need for a systematic literature review, (2) selecting data sources, (3) selecting search string, (4) defining research selection criteria, (5) defining research selection process and (6) defining data extraction and synthesis.

In favor of increasing readability, we only provide a summary of the extracted properties in this section (table~\ref{table:properties}), while the review protocol and the included stages are described in more detail in Appendix A\ref{sec_appendix}.

\begingroup
\footnotesize
\begin{longtable}{ | L{.25\textwidth} | p{.7\textwidth} | } \hline \hline
\rowcolor{black!15}\Centering Property & \Centering Description \\ \hline
P1 - Publication year & Publication year \\ \hline
P2 - Human-oriented Factors &  What human factors or concepts, if any, were analyzed or considered in the study. As a starting point, we used the list of BSE concepts defined in our previous publication~\cite{lenberg2015behavioral}. \\ \hline
P3 - Type of Organizational Change & What type of organizational change was analyzed in the study? Did the study analyze a specific type of change, e.g. agile transition or transition to global software development, or did it analyze organizational change in general. \\ \hline
P4 - Faculty affiliation & The faculty affiliation of the researcher(s) conducting the study, classified as (1) business (accounting, economics, finance, management, marketing), (2) humanities (art, history, languages, literature, music, philosophy, religion, theater), (3) natural and applied sciences (biology, chemistry, computer science, engineering, geology, mathematics, physics, medicine) or (4) social sciences (anthropology, education, geography, law, political science, psychology, sociology). \\ \hline
\caption{Properties extracted for each paper in the systematic literature review.}
\label{table:properties}
\end{longtable}
\endgroup

\subsection{Results}
In total, we scanned almost 800 papers. Of these, 85 remained after we read the titles and abstracts, and 41 were finally included as primary studies after a full review~\ref{table:No_Search_Results}.
The analysis of the primary studies is presented in the following section per extracted property (see Table~\ref{table:properties}).

\begin{figure}[ht]
\centering
\includegraphics[width=0.85\linewidth]{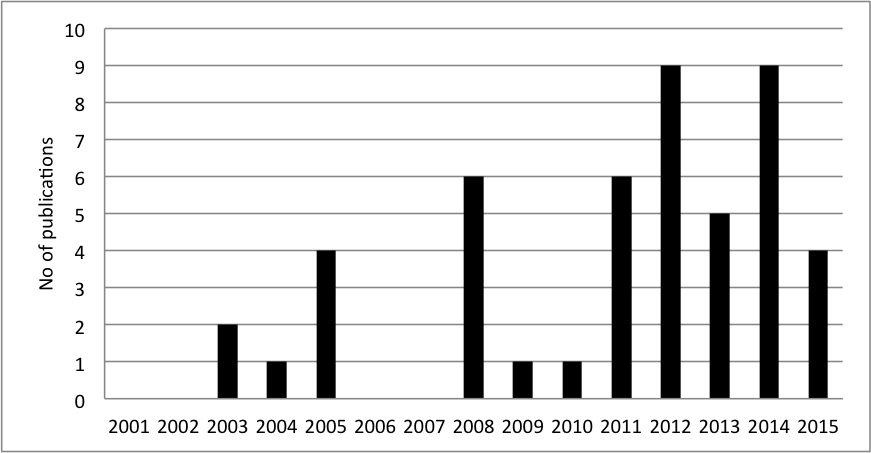}
\caption{Result for P1 - Figure showing the temporal distribution of the included publications.}
\label{fig:P1}
\end{figure}

Figure~\ref{fig:P1} presents a diagram of the temporal distribution for the included publications. It indicates an increasing interest in organizational change research, for example, more than 85\% of the studies have been published after year 2007.

\subsubsection{Property 1 - Publication Year}

\subsubsection{Property 2 - Human-oriented Factors Affecting Change Outcome}

As shown in table~\ref{table:P2}, 40\% percent of the publications did consider the human-oriented factors of organizational change. 
It should be noted that several software engineering researchers acknowledge the need to explore human-oriented factors~\cite{gandomani2015impact, nikitina2011developer, gandomani2014human, gandomani2014agile, gandomani2013important, seger2008agile, nerur2005challenges}, and, for example, Parizi, Gandomani and Nafchi~\cite{parizi2014hidden} stressed that the human-oriented factors have not be explored as a part of the change process even though agile transition is all about changing the people' mindset.
The most common human-oriented factors in the review, considered in six publications or more, were management, knowledge, participation and need for change.

The most common factor was \emph{management}. The included publications stated that the organizational changes will inevitably fail if it lacks support and proposed than management should be aware of, committed to, understand and support the organizational changes~\cite{gandomani2014agile, gandomani2014exploratory, hajjdiab2012industrial, cao2009framework}.

Yet another common factor was employee feeling of \emph{participation} in the organizational change~\cite{mohanarajah2015improved, serour2005resistance, nikitina2011developer, gandomani2015impact, nikitina2012scrum, mathiassen2005managing}. An effective ingredient in defeating peoples' resistance to change is to invite and encourage the employees to participate in the planning as well as the implementation of the change~\cite{serour2005resistance}.

Furthermore, second most frequently considered factor was employees' \emph{knowledge} about the result, i.e. the end-state, of the organizational change. The research indicated that focusing on raising the employees' knowledge with, for example, training would lead to an easier and faster transition, more motivated employees and fewer challenges during the transition process~\cite{gandomani2015impact, gandomani2014human, gandomani2014agile, nikitina2012scrum, conboy2010people, nikitina2011developer}.

Finally, the included publications stated that the employees must understand and feel that the organizational change is necessary~\cite{gandomani2015impact, gandomani2014human, serour2005resistance, seger2008agile, lenberg2015human}. The feeling of need for the change shapes the employees' perceptions about the change process~\cite{gandomani2014human}.

 \begin{table}[ht]
\footnotesize
\centering
\begin{tabular}{ | L{.35\textwidth} | C{.2\textwidth} |} \hline
\rowcolor{black!15}\Centering Human aspects & \Centering No of publications  \\ \hline
Management & 10(21\%)     \\ \hline
Participation & 7(13\%)     \\ \hline
Knowledge & 6(13\%)     \\ \hline
Need of change & 6(13\%)     \\ \hline
Organizational culture  & 5(13\%)     \\ \hline
Motivation  & 4(8\%)     \\ \hline
Resistance to change  & 3(6\%)     \\ \hline
Trust  & 3(6\%)     \\ \hline
Worry  & 3(6\%)     \\ \hline
Stress  & 2(4\%)     \\ \hline
Commitment  & 2(4\%)     \\ \hline
Miscellaneous (cognitive fit, self efficacy, psychological needs)  & 5(10\%)     \\ \hline
None  & 29(60\%)     \\ \hline
\end{tabular}
\caption{Result for P2 - Factors Affecting Change Result.}
\label{table:P2}
\end{table}

\subsubsection{Property 3 - Type of Organizational Change}

\begin{table}[ht]
\footnotesize
\centering
\begin{tabular}{ | L{.35\textwidth} | C{.2\textwidth} |} \hline
\rowcolor{black!15}\Centering Type of change & \Centering No of publications  \\ \hline
Agile & 30(63\%)     \\ \hline
General, none-specific changes & 8(17\%)     \\ \hline
Software process improvement & 5(10\%)     \\ \hline
Global software development & 2(4\%)     \\ \hline
Lean & 1(2\%)     \\ \hline
Model driven development & 1(2\%)     \\ \hline
Object orientation & 1(2\%)     \\ \hline
\end{tabular}
\caption{Result for P3 - Number of publications per type of organizational change.}
\label{table:P3}
\end{table}

As shown in table~\ref{table:P3}, the single most researched type of change was, by far, \emph{agile} transition with a total of 30 publications, of which three were literature reviews. Two literature reviews focused on identifying success and failure in agile transition~\cite{mohanarajah2015improved, razavi2014agile}, while the third review~\cite{gandomani2014agile} rated the different concepts within the agile methodology in order for the companies to better prioritize the adoption process. These reviews are, clearly, not related to the actual change process, but rather to the end-state or the aim of the change. Although the end-state has bearing on the actual organizational change process, it is only indirectly related. Hence, the agile related research has, so far, focused on \emph{what} to change rather than \emph{how} to conduct the change. 

Of the 27 none-review agile related publications, more than half (14)~\cite{gandomani2013exploring, noordeloos2012rup, prokhorenko2012skiing, nikitina2012scrum, hajjdiab2012industrial, nikitina2011developer, smits2011agile, savolainen2010transition, li2010transition, seffernick2007enabling, schatz2005primavera, seger2008agile, cohn2003introducing, olsson2012climbing} were lessons learned studies where the researchers followed a agile adoption process at one or a few companies.

Moreover, the review identified eight publications that did not study any specific organizational change; instead, they analyzed software engineering \emph{change in general} (see Table~\ref{table:P3}). Two publications showed that organizational change has a momentary negative effect on quality~\cite{sato2013effects, mockus2010organizational}. Another study showed that transformational leaders could act as change agents and, thereby, facilitate the change process~\cite{Angeline201514299}. Two publications focused on the human-oriented factors~\cite{serour2005resistance, lenberg2015human}. The first focused on resistance to change and how a general focus on the human aspects can reduce the resistance. The second publication stressed that software engineers, primarily the developers, consider organizational change as one of the most important areas of research. 

The third most frequent type of change was \emph{software process improvement} (SPI), which has been studied from a broad perspective. Several studies had an organizational focus~\cite{mathiassen2005managing, kautz2004understanding}, but there were also studies considering the individual employee perspective~\cite{lavallee2012impacts}. Regarding the latter, a review by Lavallée and Robillard summarizes the impact SPI has had on the software engineers. Positive impacts were, for example, the reduction in the number of crises, increased team communications and increased morale. Negative impacts include increased overhead on software engineers and the fact that SPI is oriented toward management and process quality, and not towards developers and product quality.

The problem of implementing SPI improvements in organizations was acknowledged in studies by Mathiassen et al.~\cite{mathiassen2005managing}, and Kautz and Nielsen~\cite{kautz2004understanding}. Mathiassen et al. stated that many SPI improvements struggle because the SPI literature have failed to communicate that change management is a key to success. Kautz and Nielsen propose a framework to help the implementation, which is focused on innovation and based on three-way process: individualist, structuralist and interactive process. The individualist perspective assumes that single individuals are the main source of change in organizations, the structuralist perspective assumes that innovation is determined by objectively existing organizational characteristics and the interactive process perspective assumes that innovation is a dynamic, continuous phenomenon of change over time in which various factors have mutual impact on each other.

Overall, the included review publications indicate that SPI research so far has been focused on identifying and evaluating what processes and practices to improve, not identifying the best way in which the actual improvements should be implemented in the organization.

\subsubsection{Property 4 - Faculty Affiliation}
\begin{table}[ht]
\footnotesize
\centering
\begin{tabular}{ | L{.35\textwidth} | C{.2\textwidth} |} \hline
\rowcolor{black!15}\Centering Faculty affiliation & \Centering No of publications  \\ \hline
Business & 0(0\%)     \\ \hline
Humanities & 0(0\%)     \\ \hline
Natural sciences & 45(94\%)     \\ \hline
Social sciences & 0(0\%)     \\ \hline
Natural sciences + business & 2(4\%)     \\ \hline
Natural sciences + social sciences & 1(2\%)     \\ \hline
\end{tabular}
\caption{Result for P4 - Number of publications per faculty affiliation.}
\label{table:P4}
\end{table}

Table~\ref{table:P4} presents the faculty affiliation of the researchers in the included publications. As can be seen, almost all, except three, of the publications have been conducted by researchers from technical faculties only, primarily computer science.

\subsection{Discussions}

The result of the review indicates an increasing interest from researchers in software engineering organizational change, which is aligned with the overall trend in the BSE research area~\cite{lenberg2015behavioral}. The single most studied type of organizational change for the past fifteen year has, by far, been transition to an agile methodology. However, other types of organizational changes have also been explored, for example software process improvement, transition to global software development and transition to model driven development. Furthermore, several publications emphasize the importance of human factors in organizational change, and the most frequently considered human factors among the included studies were; management, knowledge, need for change and employee participation. In addition, the review indicates that the researchers so far have been focused on \emph{what} to change, i.e. in describing the desired end-state, rather than exploring \emph{how} the actual organizational change shall be conducted in order to be successful and reach the desired goal. Researchers from technical faculties only, primarily computer science, have conducted almost all, except three, of the included studies.

More specifically, the review indicates that some software researchers do not consider process transformation in software engineering, e.g. agile transformation, to be an organizational change. For example, the term 'organizational change' is only used in 19 of the 48 included publications, i.e. in less than 40\%. In addition, standard change models in work- and organizational psychology, such as  Kotter~\cite{kotter2002heart}, were only used in one of the included publications~\cite{smits2011agile}. From a work- and organizational perspective, it is rather clear that an agile transition should be considered an organizational change. Although, it should be noted the organizational researchers disagree on the definition~\cite{van2005alternative} of organizational change, but it is usually considered to be approach in an organization for ensuring that changes are smoothly and successfully implemented to achieve lasting benefits.

Not to recognize process transformation in software engineering as an organizational change might have some drawbacks. Since organizational change has been studied within work- and organizational psychology for several decades, software engineering researchers risk reinventing the wheal instead of adapting or building on existing theories. One reason that process transformation is not considered an organizational change might be that almost all software engineering change research are conducted by researchers from technological faculties.
approaches and methods.

Moreover, it is encouraging that 40\% of the publications in this review have considered the human factors. Although, this does not mean that these concepts were actually studied in depth. Many of the publications acknowledged that the human factors are important~\cite{gandomani2015impact, nikitina2011developer, gandomani2014human, gandomani2014agile, gandomani2013important, seger2008agile, nerur2005challenges}, however, the human factors are seldom a core part of the studies. Instead, they are often used as generalized terms that are not operationalized into more specific constructs that could be analyzed more thorough~\cite{serour2007radar, gandomani2013obstacles, gandomani2013important, nikitina2011developer}. This too might be a consequence of the faculty homogeneity mention earlier. It is not surprising that researchers from technical faculties have emphasized the technical and process oriented factors of organizational change. Naturally, researchers tend to conduct research related to their area of expertise. Thus, a deeper understanding of the human aspects of organizational change in software engineering calls for a more interdisciplinary approach, meaning that more studies should be performed in collaboration by researchers from technological and social science faculties.

In the work- and organizational psychology community, there is an ongoing discussion regarding the nature of change~\cite{bouckenooghe2010positioning}, which describes how changes emerge and evolve over time~\cite{porras1991organization}. The researchers distinguish between two different facets: episodic and continuous change. The episodic change, also called programmatic change, is intentional and assumes that organizations are driven by a clearly defined purpose and that the change has a clearly defined goal that will contribute to that purpose~\cite{tenkasi2003social}. The continuous change, on the other hand, is ongoing, evolving and cumulative~\cite{orlikowski1996improvising}. This review indicates that software engineering researchers have not focused on the change process in itself. Instead, they have put a lot of efforts into describing a clear end-state and defining \emph{what} the organizational change should accomplish, which is typical for episodic change. The software engineering researchers should be aware of that there are prominent researchers that do not promote this type of change, for example Beer, Eisenstat and Spector~\cite{Beer1990158} in one of the most influential articles written as early as 1990.

Finally, we recognize and understand the importance of conducting research related to \emph{what} to change; however, we believe that the research needs to be balanced and that future research would benefit putting more effort into researching \emph{how} to conduct the organizational changes. This is also acknowledged by other software engineering researcher, for example Nikitina, Kajko-Mattsson and Stråle~\cite{nikitina2012scrum} and Serour and Winder~\cite{serour2007radar}.

%
%
\section{Software Engineers' Attitudes Towards Organizational Change - a Case Study } %
Previous research has shown that employees' attitudes play a vital part in the organizational change process. It is, primarily from an industrial but also from a research perspective, of interest to understand what underlying constructs that affect them. Therefore, the purpose of this study was to create, verify and validate a first-order model that predicts software engineers' attitude towards organizational change.

\subsection{Method} %
\label{sec_methods_survey}
In a previous publication~\cite{lenberg2015behavioral}, we have argued that behavioral software engineering (BSE) research, i.e. human-aspect related research in software engineering, would benefit from becoming more interdisciplinary. The social sciences have over one hundred years of experience in the study of behavior, and their gained knowledge in, for example, research design and methodology could be used to leverage BSE research. We acknowledge, on the other hand, that software engineering is a highly complex activity and that software engineering researchers' domain knowledge is imperative.

Thus, in order to create the model, we used a method that has been proven within social science research, the \emph{traditional measurement development procedure}~\cite{viswanathan2005measurement}. The procedure included the following five steps: (1) model domain definition and delimitation, (2) item generation and questionnaire design, (3) data collection procedure (4) factor analysis, (5) internal consistency and (6) data analysis.

\subsubsection{Model Domain Definition and Delimitation}
We chose to use a quantitative research design with questionnaires. According to Creswell \cite{creswell2013research}, if the problem is identifying factors that influence an outcome, the utility of an intervention, or understanding the best predictors of outcomes, then a quantitative approach is preferred.

A limitation in this study was the number of questions we were allowed by the company to use in the questionnaire, which restricted the study to include five variables. We acknowledge that such first-order models cannot be considered complete, rather, they are to be recognized as a first approximations that captures the most significant effects.

Regarding the dependent variables, i.e. attitude towards organizational change, we aimed to use constructs that had already been verified in previous research. In a literature review, Myungweon~\cite{choi2011employees} identified that the constructs most frequently used to measure such attitudes were \emph{openness to change}, \emph{readiness for change}, \emph{commitment to change} and \emph{cynicism about organizational change}. These four constructs are similar in that they all reflect an individual’s overall positive or negative judgment of a specific change initiative. 

Of these four constructs, we chose to use \emph{readiness for change} and \emph{openness to change} for the following reasons. We chose \emph{readiness for change} since this was the most commonly used construct according to the literature review. In addition, the organizational change we explored in the study was planned by the management and, according to Miller~\cite{miller1994antecedents}, \emph{openness to change} is defined as an initial condition for such planned change. Finally, since specific attitudes are a better predictor of behavior than general~\cite{ajzen1977attitude}, we wanted the questions measuring the attitudes to be directed towards the specific organizational change, not to organizational changes in general. The \emph{readiness for change} and \emph{openness to change} constructs meet this requirement. 

Furthermore, the process we used to identify independent variables with a significant effect on organizational change in software engineering organizations included three steps. First (step 1), we conducted a systematic literature review (SLR) in order to get a better understanding of the domain and, also, to identify factors that software engineering researchers, so far, have deemed the most important. Second (step 2), since our ambition was to combine the software engineering domain knowledge, gained through the SLR, with knowledge in organizational psychology, we examined if these identified factors have been used in existing organizational change theories. Third (step 3), considering that it was important that our model was useful in practice, we made a preliminary assessment if the factors were susceptible to situational differences, i.e. that they could change over time as individuals’ experiences change. In other words, we sought to include factors that were better conceptualized as states than as personality traits and thus could be affected by the managers.

\paragraph{Step 1} The SLR identified \emph{management} as the most frequently occurring human-oriented factor; however, its definition was rather broad and diverse, meaning that it referred to the both leadership style of the individual manager but also to the commitment, knowledge and support from management in general. We did not find any appropriate way to operationalize such a diverse factor without jeopardizing the validity and reliability of the measurement and, consequently, we decided not to include it into our model.

Apart from \emph{management}, the most frequently considered factors were \emph{knowledge}, \emph{need for change} and \emph{participation}. \emph{Knowledge} relates to the software engineer's understanding about the outcome of the planned change. If he/she has sufficient insights in order to determine how it will affect the company, but also how it will affect his/hers everyday work. In addition, this factor also relates to if the software engineer has previous experience of the change outcome. The second factor (\emph{need for change}) describes if the employee understands why the change is necessary, i.e. to the reasons behind the initiation of the change. Does the software engineer feel that the organization needs to change, or does he/she think that everything is working quite well? The third factor (\emph{participation}) relates to the software engineer's feeling of influence over the change process.

\paragraph{Step 2} We made the assessment that the three identified factors, or at least factors with similar meaning to them, have previously been used in organizational change theories. Rendahl et al.~\cite{rendahl1996att} have proposed a theoretical model claiming that an effective organizational change is the sum of the three constructs: (1) that the employees understand the causes of the change, (2) that the employees accepts the proposed solution, and (3) that there is quality in the solution. Our hypothesis is that the \emph{need for change} factor in our study is related to the cause of the change (1) and the \emph{knowledge} factor is a prerequisite to the acceptance of the solution (2). In addition, according to Rendahl, a key component in achieving a quality solution (3) is to involve the employees in the change process, which means that it relates to the \emph{participation} factor in our study.

The \emph{need for change} factor is, also, related to the first step in Kotter's~\cite{kotter1996transformation} eight-step change process, i.e. \emph{sense of urgency}. Kotter argues that the purpose of this first step is to help others see the need for change and the importance of acting immediately. By developing a sense of urgency around the need for change, you will be able to spark the initial motivation to get things moving.

Furthermore, regarding \emph{knowledge}, Beer et al.~\cite{Beer1990158} and Robey et al.~\cite{robey2002learning} have shown that effective training strengthens employees' commitment to change.

\paragraph{Step 3} Finally, we made the assessment that the three factors are to be conceptualized as states, over which managers exert influence. For example, we made the assumptions that \emph{knowledge} could be affected by training, that \emph{need for change} could be affected by adequate information and, finally, that \emph{participation} could be affected by the design of the change process.

Hence, as shown in figure~\ref{fig:model}, we chose to include the variables \emph{knowledge}, \emph{participation}, \emph{need for change}, \emph{openness to change} and \emph{readiness for change} into our model.

\subsubsection{Data Collection Procedure} %
\label{sec_methods_survey_participants}
An overview of the company and a brief description of the organizational change is found in the background section (\ref{sec_background_company}). The employees were working at three different sites in Sweden. Two of the sites were roughly the same size in terms of number of employees (20), while the third site was approximately half the size. The employees had different roles and responsibilities and worked as software developers, project managers, team leaders or managers. The majority of the employees (90\%) had a master degree in engineering, either in software engineering, engineering physics or electronics.

The questionnaires were distributed to the employees at a biweekly department meeting in October of 2014, one week after the management had announced the organizational change. Before filling out the survey the participants were informed about the purpose of the study; that it was anonymous and that it was voluntary to participate. They were also informed that the researcher would not share the raw data with other researchers nor the management, and that no attempts would be made to identify who answered which questionnaire. During the meeting, it became clear that the employees were interested in the outcome of the study and, therefore, they were promised feedback from the researcher after the survey was completed and the data was analyzed.

Of the 65 employees at the department, 57 choose to participate in the study giving a response rate of 88\%. Approximately one per cent of the respondents were below the age of 25, and approximately 10\% were older than 45 years. The department had been growing for the last ten years, and about 40\% of the respondents had been working for the department for five year or less, while about 15\% had been working there for more than 15 years. About 55\% of the respondents had more than 10 years of IT work experience, 33\% between six and ten years experience and 12\% fewer than six years. Approximately 80\% of the respondents were male.

\subsubsection{Item Generation and Questionnaire Design}
\label{sec_methods_survey_instrument}
The questionnaire included two main parts. The first part included five items (or questions) related to background information about the respondent, The respondents were asked what type of work they performed (administrative or technical), if he/she was responsible for personnel or not, for how many year they had been working (0-5 years, 6-10, 11-15, \textgreater15) and, finally, for how many years that had been working for the department (0-5 years, 6-10, 11-15, \textgreater15).
The second part included items related to the four variables in the model, see figure~\ref{table:items_model}. The answers to all of these questions were measured using a five-point Likert scale with the following alternatives: 1 "strongly agree", 2 "agree", 3 "neither agree nor disagree", 4 "disagree" and 5 "strongly disagree".

\emph{Openness to change} was measured using an eight-item measure developed by Miller, Johnson and Grau~\cite{miller1994antecedents}, and is conceptualized as willingness to accommodate and accept change, positive affect about the potential consequences of the change, and considered a necessary, initial condition for successful planned change.
\emph{Readiness for change} was measured using six items proposed by Cunningham et al.~\cite{cunningham2002readiness}, and included questions reflecting the precontemplative, contemplative, preparatory, action and maintenance stages of the change model developed by Prochaska et al.~\cite{prochaska1994stages}.
The \emph{participation} variable was assessed using four items developed by Wanberg et al.~\cite{wanberg2000predictors}, which measured the extent to which employees perceived that they had input into the change.
We updated these two variables and directed the question towards the specific organizational change that the company was currently undergoing, i.e. transition to team-based development.
Although we aimed to use constructs that had already been verified in previous research, we did find any appropriate for the final two variables, i.e. \emph{knowledge} and \emph{need for change}. Consequently, for these, we compiled the items ourselves using the constructs definitions, which were based on information gathered in the literature review.
As stated previously, the \emph{knowledge} variable was relates to the employees knowledge about the outcome of the change. If he/she has sufficient in-sights regarding the change in order to see how it will affect the company, but also how it will affect his/hers everyday work. This factor also holds a component related to the experience of the outcome. We used four items to capture these aspects, see table~\ref{table:items_model}.
Finally, four items were used to compile the \emph{need of change} variable. The first three were related to the software engineers' opinion of the departments current way-of-working and to the quality of the systems that department produced, while the forth question was more directly related to the need for change.

\begingroup
\footnotesize
\begin{longtable}{| L{.05\textwidth} | L{.20\textwidth}| L{.60\textwidth} | C{.05\textwidth} |} \hline
\rowcolor{black!15} Identifier & Variable & \Centering Question  & \Centering Reversed \\ \hline

O1 & Openness to change &I would consider myself to be "open" to the changes that team based development will bring to my work role. & \\ \hline
O2 & Openness to change &Right now, I am somewhat resistant to the proposed changes in work teams. & (R) \\ \hline
O3 & Openness to change &I am looking forward to the changes in my work role brought about by the implementation team based development. & \\ \hline
O4 & Openness to change &In light of the proposed changes regarding team based development, I am quite reluctant to consider changing the way I now do my work. & (R) \\ \hline
O5 & Openness to change &I think that the implementation of team based development will have a positive effect on how I accomplish my work. & \\ \hline
O6 & Openness to change &From my perspective, the proposed changes regarding team based development will be for the better. & \\ \hline
O7 & Openness to change &The proposed changes regarding team based development will be for the worse in terms of the way that I have to get my work done. & (R) \\ \hline
O8 & Openness to change &I think that the proposed changes in the work teams will have a negative effect on how I perform my role in the organization. & (R) \\ \hline

R1 & Readiness for change &The programme or area in which I work functions well and does not have any aspects which need changing. & \\ \hline
R2 & Readiness for change &There's nothing that I really need to change about the way I do my job to be more efficient. &  \\ \hline
R3 & Readiness for change &I've been thinking that I might want to help change something about the programme or area in which I work. & (R) \\ \hline
R4 & Readiness for change &I plan to be involved in changing the programme or area in which I work. & (R) \\ \hline
R5 & Readiness for change &I am working hard to help improve aspects of the programme or area in which I work. & (R) \\ \hline
R6 & Readiness for change &We are trying to make sure we keep changes/improvements my programme/area has made. & \\ \hline

P1 & Participation & I have been able to ask questions about the changes regarding team-based development. & \\ \hline
P2 & Participation & I have been able to participate in the implementation of team-based development. & \\ \hline
P3 & Participation & I have some control over the changes regarding team-based development. & \\ \hline
P4 & Participation & If I wanted to, I could have input into the decisions being made about team based development. & \\ \hline

K1 & Knowledge & I have sufficient knowledge about team-based development in order to determine how that will affect my work. & \\ \hline
K2 & Knowledge & I believe that I have a good knowledge about team-based development. & \\ \hline
K3 & Knowledge & I have experience in team-based development. & \\ \hline
K4 & Knowledge & I can determine if my tasks are suited to be performed using a team-based development. & \\ \hline

N1 & Need for change & I think that this department's current way-of-working is cost effective. & (R) \\ \hline
N2 & Need for change & I think that this department develops systems with the appropriate quality. & (R) \\ \hline
N3 & Need for change & The documents that this department delivers to the customers hold an appropriate quality. & (R) \\ \hline
N4 & Need for change & I think that this department needs to change its way-of-work. & (R) \\ \hline
\caption{The table shows the questions used to compile the four variable in the model. The last column indicates whether the response is to be inverted when creating the index variables.}
\label{table:items_model}
\end{longtable}
\endgroup

\subsubsection{Factor analysis}
Confirmatory factor analysis (CFA) was used to test whether measures of the items are consistent with our understanding of that item's nature, and, as such, to test whether the data fit a hypothesized measurement model~\cite{fabrigar1999evaluating}.
The sample size is important in such analysis and several guiding rules of thumb are cited in the literature; however, Sapnas and Zeller state that 50 cases is adequate~\cite{sapnas2002minimizing}. In addition, the Kaiser-Meyer-Olkin measure of sampling adequacy was 0.74 in our analysis, and a value above 0.50 is considered suitable~\cite{hair2006multivariate}.
The result of the CFA are presented in table~\ref{table:CFA}. As shown, four items (O1, R6, K4 and N4) were dropped on the basis of the CFA and not included in the subsequent analysis.

We have chosen CFA for our analysis since it is a well-established and broadly used method in social science and psychological research. Factor analysis is sometimes criticized~\cite{Fabrigar1999,Eijk2015} and several alternatives have been proposed~\cite{Lee1981,Flora2004,Muthen2012,Conti2014}. However, we note that the criticism primarily focuses on exploratory factor analysis rather than the confirmatory factor analysis that we have employed. Even though these two methods have similarities they are both conceptually and statistically different.

Still, Flora et al~\cite{Flora2004} showed how CFA performed adequately only for large sample sizes even though it was robust to modest violations of its assumptions of normally distributed, continuous, latent model variables. In their simulations Robust Weighted Least Squares regression showed superior performance. Bayesian alternatives to CFA was proposed already in the early 1980's~\cite{Lee1981} and allow for more flexible specification of models and less restrictive constraints on latent variables. For example, a recent approach of Conti et al~\cite{Conti2014} infers the number of factors along with other parameters and can utilize all available information without discarding measurements. In particular, Muthén and Asparouhov's~\cite{Muthen2012} Bayesian SEM approach `[replaces] the parameter specification of exact zeros and exact equalities with approximate zeros and equalities` and can thus better represent substantive theories without multiple, often ad hoc, model rejection and retry steps as can happen when using CFA. In future work, we can consider analysing our data with one or more of these alternative methods. However, since we tested for normality and used state-of-the-art model fitting assessment procedures we do not consider the choice of CFA as a threat to the validity of our results.

\begin{table}[H]
\footnotesize
\centering
\begin{tabular}{| L{.10\textwidth} | C{.10\textwidth} | C{.10\textwidth} | C{.10\textwidth} | C{.10\textwidth} | C{.10\textwidth} |} \hline
\rowcolor{black!15} Item & \Centering Readiness for change & \Centering Openness to change & \Centering Knowledge & \Centering Need for change & \Centering Participation \\ \hline
O2 & .74    &       &       &       & \\ \hline
O3 & .80    &       &       &       & \\ \hline
O4 & .58    &       &       &       & \\ \hline
O5 & .72    &       &       &       & \\ \hline
O6 & .80    &       &       &       & \\ \hline
O7 & .84    &       &       &       & \\ \hline
O8 & .85    &       &       &       & \\ \hline
R1 &        &   .51 &       &       & \\ \hline
R2 &        &   .53 &       &       & \\ \hline
R3 &        &   .80 &       &       & \\ \hline
R4 &        &   .74 &       &       & \\ \hline
R5 &        &   .64 &       &       & \\ \hline
K1 &        &       & .81   &       & \\ \hline
K2 &        &       & .91   &       & \\ \hline
K3 &        &       & .73   &       & \\ \hline
N1 &        &       &       & .54   & \\ \hline
N2 &        &       &       & .83   & \\ \hline
N3 &        &       &       & .70   & \\ \hline
P1 &        &       &       &       & .77  \\ \hline
P2 &        &       &       &       & .85 \\ \hline
P3 &        &       &       &       & .80 \\ \hline
P4 &        &       &       &       & .77 \\ \hline
\end{tabular}
\caption{Principal component analysis, varimax rotation with Kaiser normalisation. For display purposes, loads less than 0.4 are suppressed. The first column refers to the items (questions) in table ~\ref{table:items_model}.}
\label{table:CFA}
\end{table}

\subsubsection{Internal Consistency}
Internal consistency is the degree to which every item measures the same construct. Cronbach's alpha, a statistical measurement calculated from the pairwise correlations between the items, was used as a lower-bound estimate of the internal consistency. The estimates are shown in table~\ref{table:Cronbach}. As can be seen, the \emph{need for change} variable has a somewhat lower value compared to the other variable; however, Streiner~\cite{streiner2003starting} and Nunnally~\cite{nunnally1967psychometric} suggested that a minimum alpha of 0.6 sufficed for early stage of research.

\begin{table}[H]
\footnotesize
\centering
\begin{tabular}{| L{.20\textwidth} | C{.15\textwidth} |} \hline
\rowcolor{black!15} Variable & \Centering Cronbach Value \\ \hline
Openness to change      &   .92 \\ \hline
Readiness for change    &   .71 \\ \hline
Knowledge               &   .85 \\ \hline
Need for change         &   .68 \\ \hline
Participation           &   .86 \\ \hline
\end{tabular}
\caption{The Cronbach's alpha value for the included variables. Streiner~\cite{streiner2003starting} and Nunnally~\cite{nunnally1967psychometric} recommended .50 to .60 for the early stages of research, .80 for basic research tools, and .90 as the for clinical purposes. }
\label{table:Cronbach}
\end{table}

\subsubsection{Data Analysis}
In order to test the proposed model, multiple linear regression analysis was used with \emph{openness to change} as dependent variable, while \emph{knowledge}, \emph{need for change} and \emph{participation} were independent variables. The analysis was based in the procedures outlined by Meyers~\cite{meyers2006applied}.

Before conducting the analysis, we verified that the collected data actually could be analyzed using linear regression. A visual analysis of scatter plots for all variables indicated a linear relationship. Further, we checked the homoscedasticity and normality of residuals with the Q-Q-Plot. The plot indicated that in our multiple linear regression analysis there is no tendency in the error terms. Regarding autocorrelation, the Durbin-Watson value for the model was d = 2.215, which is between the two critical values of 1.5 $<$ d $<$ 2.5 and, therefore, we can assume that there is no first order linear auto-correlation in our multiple linear regression data. Finally, the data was analyzed in order to determine the presence of multicollinearity. The variance inflation factors~\cite{kleinbaum1998applied} were all well below three (maximum VIF was 1.13), indicating a small risk for multicollinearity.

\subsection{Results} %

\begin{table}[ht]
\footnotesize
\centering
\begin{tabular}{| C{.20\textwidth} | C{.15\textwidth} | C{.15\textwidth} | C{.15\textwidth} | C{.15\textwidth} |} \hline
\rowcolor{black!15}\Centering Variable & \Centering Openness to change & \Centering Knowledge & \Centering Need for change & \Centering Participation \\ \hline
Readiness for change    &   .334*    & .085      & .314*    & .243** \\ \hline
Openness to change      &   -       &   .449*    & .247**    & .339* \\ \hline
Knowledge               &   -       &   -       & .033    & .054 \\ \hline
Need to change          &   -       &   -       & -       & -.108 \\ \hline
\end{tabular}
\caption{Pearson r correlations of the variables. '*' means that the correlation is significant at 0.05 level. '**' means that the correlation is significant at 0.10 level.}
\label{table:Correlation}
\end{table}

\begin{table}[ht]
\footnotesize
\centering
\begin{tabular}{| C{.15\textwidth} | C{.1\textwidth} | C{.1\textwidth} | C{.1\textwidth} | C{.1\textwidth} |C{.1\textwidth} |C{.15\textwidth} |} \hline
\rowcolor{black!15}\Centering Model & \Centering B & \Centering SE-b & \Centering Beta & \Centering Pearson r & \Centering sr\textsuperscript{2} & \Centering Sig\\ \hline
(Constant)      &   .829   & .500  &       &       &       & .103 \\ \hline
Knowledge       &   .349    & .068  & .522  & .449  & 0.271 & .000 \\ \hline 
Participation   &   .265    & .080  & .340  & .339  & 0.114 & .002 \\ \hline
Need for change  &   .277    & .106  & .266  & .247  & 0.070 & .012 \\ \hline
\end{tabular}
\caption{The raw and standardized regression coefficients of the predictors together with their correlations with \emph{openness to change}, their squared semi-partial correlations (sr\textsuperscript{2}) and the significance level. The dependent variable was \emph{openness to change}.}
\label{table:Regression_opennesstochange}
\end{table}

\begin{table}[ht]
\footnotesize
\centering
\begin{tabular}{| C{.15\textwidth} | C{.1\textwidth} | C{.1\textwidth} | C{.1\textwidth} | C{.1\textwidth} |C{.1\textwidth} |C{.15\textwidth} |} \hline
\rowcolor{black!15}\Centering Model & \Centering B & \Centering SE-b & \Centering Beta & \Centering Pearson r & \Centering sr\textsuperscript{2} & \Centering Sig\\ \hline
(Constant)      &   2.233   & .450  &       &       &       & .000 \\ \hline
Knowledge       &   .029    & .061  & .059  & .085  & 0.003 & .641 \\ \hline 
Participation   &   .258    & .095  & .342  & .314  & 0.116 & .009 \\ \hline
Need for change &   .156    & .072  & .277  & .243  & 0.076 & .033 \\ \hline
\end{tabular}
\caption{The raw and standardized regression coefficients of the predictors together with their correlations with \emph{readiness for change}, their squared semi-partial correlations (sr\textsuperscript{2}) and the significance level. The dependent variable was \emph{readiness for change}.}
\label{table:Regression_readinessforchange}
\end{table}

\emph{Knowledge}, \emph{need for change} and \emph{participation} were used in two separate standard multiple linear regression analysis to predict \emph{openness to change} (OTC) and \emph{readiness for change} (RFC). The correlations of the variables, shown in Table \ref{table:Correlation}, were overall low with a maximum value of .45 for \emph{openness to change} and \emph{knowledge}.

The analysis showed that the models for OTC and RFC are statistically significant (F(3, 52) = 15.25, p \textless .001; F(3, 52) = 3.82, p = .015). The former accounts for 44\% (R\textsuperscript{2} = .468, Adjusted R\textsuperscript{2} = .437) of the variance, while the latter accounts for considerable less, 14\% (R\textsuperscript{2} = .180, Adjusted R\textsuperscript{2} = .136). The explained variance of the OTC model is rather high compared to other studies in social science. This is, to a certain extent, the result of the relatively large number of participants in relation to a small number of variables; however, it also adds support that our hypothesized model is a good first-order approximation and captures important factors.

The regression results for the models are presented in table~\ref{table:Regression_opennesstochange} and~\ref{table:Regression_readinessforchange}. As can be seen, \emph{knowledge} is not significant in the RFC model, which somewhat clarifies the differences in explained variance between the models. However, the \emph{the need for change} and the \emph{participation} variables have the same order of magnitude, both in terms of their contribution to the models (Beta) and also in terms of the unique variance they explain, indexed by the squared semi-partial correlation in column sr\textsuperscript{2}.

\begin{figure}[H]
\centering
\includegraphics[width=0.70\linewidth]{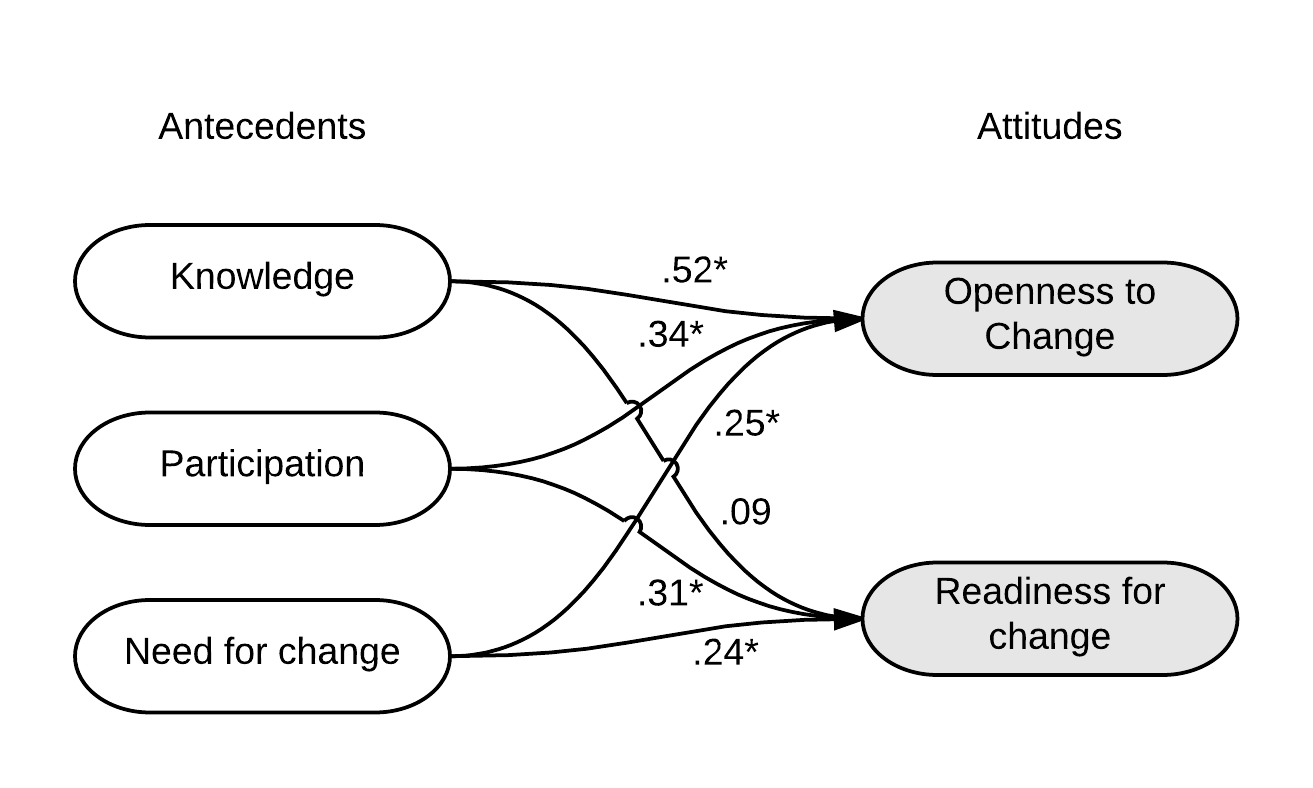}
\caption{Model used to predict \emph{openness to change} and \emph{readiness for change}. The connection line figures are the beta value from table~\ref{table:Regression_opennesstochange} and \ref{table:Regression_readinessforchange}. '*' means that the value was statistically significant at .05. }
\label{fig:model}
\end{figure}

The models, which are presented in figure~\ref{fig:model}, show that \emph{knowledge} has nearly fifty percent more impact compared to \emph{participation} (only valid for \emph{openness to change}), and that \emph{participation}, in turn, has nearly fifty percent more impact compared to \emph{need for change}.

\subsection{Discussion}
Previous research has identified employees' attitude as one of the most critical factors in organizational change. In this case study, we have used industry data to examine if the \emph{knowledge} about the intended change outcome, the understanding of the \emph{need for change}, and the feelings of \emph{participation} affect software engineers' \emph{openness to change} and \emph{readiness for change}, two commonly used attitude constructs. The result of two separate multiple regression analysis showed that \emph{openness to change} is predicted by all three concepts, while \emph{readiness for change} is predicted by \emph{need for change} and \emph{participation}. In addition, our research also provides a hierarchy with respect to the three predictive constructs' degree of impact. It shows that \emph{knowledge} has nearly fifty percent more impact compared to \emph{participation} (only valid for \emph{openness to change}), and that \emph{participation}, in turn, has nearly fifty percent more impact compared to \emph{need for change}.

Although the analysis shows that both the proposed models are statistically significant, it also indicates that the three chosen constructs capture the effects of \emph{openness to change} better than of \emph{readiness for change}. The \emph{knowledge} construct showed no significant effect for \emph{readiness for change} and the overall explained variance was considerable lower. That \emph{knowledge} about the intended change outcome is a vital component in \emph{openness to change}, whereas it seems to be neglectable in \emph{readiness for change}, has not been shown in previous research. Although, this finding provides some support to previous research~\cite{choi2011employees} that the two attitudes constructs represent different aspects of employees’ attitudes toward organizational change and the lack of one attitude does not simply represent the lack of another. This also implies that researchers, in order to understand employees' attitude towards organizational change, need to measure several different constructs. 
 
Our result also supports previous theoretical research, e.g. Rendahl's model for effective organizational change~\cite{rendahl1996att} and Kotter's organizational change process~\cite{kotter1996transformation} mentioned earlier. Alike, the result provides some support to previous research stating that when employees are well informed about the goings-on within the organization and feel included in the task and the social information network, they are likely to be open to change~\cite{miller1994antecedents, wanberg2000predictors,erturk2008trust}. However, to the best of our knowledge, there exist no previous research that has explored \emph{knowledge} in relation to \emph{openness to change}. Other attitude constructs have been studied in relation to knowledge. For example, Beer et al.~\cite{Beer1990158} and Robey et al.~\cite{robey2002learning} have shown that effective training strengthens employees' \emph{commitment to change}, and Lau and Herbert~\cite{lau2001experiences} have shown that training is the number one factor for positive staff reactions and commitment during large scale implementations.

The result has implications for practitioners. Managers can use it to increase their chances of successfully implementing change initiatives that actually result in a change in organizational behavior. First, at least in order to affect \emph{openness to change}, they should focus on increasing software engineers' \emph{knowledge} about the change. It is important to make sure that the engineers understand how the change will affect them personally in their daily work.

Also, the change process should be designed so that the software engineers have the opportunity to \emph{participate} in and influence the change. The more influence the software engineers’ feel that they have, the more open to the change they will become.

Finally, managers should try to influence the software engineers' \emph{need for change}, for example, by providing adequate information. Miller et  al.~\cite{miller1994antecedents} have shown that the perceived quality of received information about change, rather than the content of the message itself, could affect employees’ attitudes. However, according to the HSM model (described in the background section~\ref{sec_background}), the importance of the actual message content is, to a certain extent, determined by the employees' cognitive ability. With a limited cognitive ability, e.g. due to stress or high work-load, the employees might be unable to process the message content and, therefore, the message source and the quality of the information will play a more important part in the attitude formation. However, it should also be noted that such heuristic attitudes are less stable and less likely to influence behaviour, compared to those formed by systematic processing.

Further, there are several limitations in this study. We have made the assumption that there is a correlation between attitude and behavior, and that a positive attitude towards organizational change ultimately will lead to a successful organizational change. This assumption needs to be verified. In addition, due to practical limitations, this study only included three independent variables and two dependent variable. However, the attitude constructs we chose had been verified in previous research, and the process we used to select the independent variables was rigorous. Furthermore, the study was conducted in a single company and included rather few respondents. The number of respondents was, of course, limited by the size of the department, but the response rate of almost 90\% is more than acceptable.

Further research is needed in order to understand the relation between attitude towards organizational change, behavior and the success of an organizational change. The proposed first-order models can easily be extended to include more dependent attitude variables as well as additional independent variables. We also believe that there are factors unique to software engineering that affect the organizational change and the researchers therefore need to take into account. For example, software engineers often work together in teams and their behaviors are therefore, to a certain extent, controlled by the team norms. Drawing on social identity and social categorization theories, Terry et al.~\cite{terry2000attitude} argue that only the attitudes supported by in-group norms predict behavior. This implies that in order to accomplish a behavior change among software engineers, management needs to analyze if the changes are aligned with the group norms. If not, they need to find ways to alter them, otherwise the team members will not behave according to the attitudes and, in practice, no organizational change will occur.

Finally, change in software engineering organizations is a highly complex process where profound knowledge about human behavior as well as domain knowledge regarding unique software engineering factors are needed. Therefore, we conclude that an interdisciplinary research collaboration is a key element in process of finding comprehensive solutions.

\section{Conclusions}
The result shows that \emph{openness to change} can be predicted by \emph{knowledge}, \emph{need for change} and \emph{participation}, while \emph{readiness for change} can be predicted by \emph{need for change} and \emph{participation}. Our research also provides a hierarchy with respect to their degree of impact. It shows that software engineers' \emph{knowledge} about the outcome of the change has nearly fifty percent more impact compared to their understanding of the \emph{participation} (only valid for \emph{openness to change}), and that \emph{participation}, in turn, has nearly fifty percent more impact compared to their feeling of \emph{need for change} in the process. 

The result has implications for practitioners. Managers can use it to increase their chances of successfully implementing change initiatives that actually result in a change in organizational behavior. At least for \emph{openness to change}, they should focus on increasing software engineers' \emph{knowledge} about the change. It is important to make sure that the engineers understand how the change will affect them personally in their daily work.

We acknowledge that the first-order models we propose can not be considered complete, rather, it is to be recognized as a first approximation that captures the most significant effects. We believe that there are software engineering unique factors affecting the organizational change that researchers need to take into account. For example, software engineers often work in teams and their behaviors are controlled by the team norms. In order to accomplish a behavior change among software engineers, management needs to analyze if the changes are aligned with the group norms. If not, they need to find ways to alter them, otherwise the team members will not behave according to the attitudes and, in practice, no organizational change will occur.

Finally, we conclude that change in software engineering organizations is a highly complex process where profound knowledge about human behaviors as well as domain knowledge regarding unique software engineering factors are needed.

\section{Acknowledgments}
We acknowledge the support of Swedish Armed Forces, Swedish Defence Materiel Administration and Swedish Governmental Agency for Innovation Systems (VINNOVA) in the project `Aligning Requirements and Veriﬁcation Practices in Air Traffic Control Systems' (project number 2013-01199).

\section*{References}
\bibliography{OrgChange}

\section{Appendix A - Review Protocol}
\label{sec_appendix}

\begin{figure}[H]
\centering
\includegraphics[width=1\linewidth]{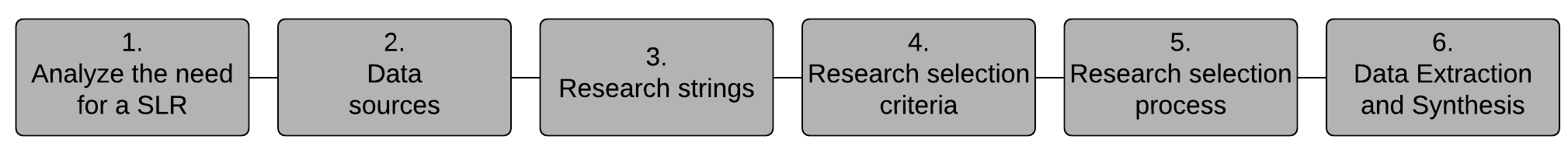}
\caption{Overview of the steps in the systematic literature review.}
\label{fig:method}
\end{figure}

The SLR was based on the guidelines described by Kitchenham~\cite{kitchenham2004procedures}. In order to reduce the possibility of researcher bias, a pre-defined review protocol was produced. The protocol described the review process, which included the following stages (also shown in Figure~\ref{fig:method}); (1) analyzing the need for a systematic literature review, (2) selecting data sources, (3) selecting search string, (4) defining research selection criteria, (5) defining research selection process and (6) defining data extraction and synthesis. These stages are described in more detail in the following sections.

\subsection{Analyzing the need for a systematic literature review}
\label{sec_methods_the_need}
For a detailed motivation, we refer to the reasoning in the introduction~\ref{sec_introduction} section above. There exists reviews that consider and analyze different aspects and concepts of organizational change in software engineering, e.g. agile development~\cite{mohanarajah2015improved, razavi2014agile,eck2014fit} and software process improvement~\cite{unterkalmsteiner2012evaluation,lavallee2012impacts,pino2008software}. However, to the best of our knowledge there exists no review covering organizational change in software engineering as a whole, which further motivates the need to complement existing systematic reviews with a broader overview.

\subsection{Select Data Sources}
\label{sec_methods_data_sources}
We consider organizational change in software engineering to be an interdisciplinary research subject. Therefore, we selected databases likely to cover both technical as well as social research; IEEE Xplore Digital library, PsycINFO, Scopus and Web of Science. The two latter are multidisciplinary databases based on abstracted and citation data and from a number of primary databases~\cite{score2009web}.

\subsection{Search String}
\label{sec_search_string}
The search string shall identify publication that include both organizational change and software engineering and, consequently, the search string combined synonyms to ‘organizational change’ with synonyms to ‘system engineers’ (defined by Cruz et al. ~\cite{cruz2011personality}) with the logical AND operator. One main contributor to organizational changes in software engineering organizations the past 15 years has been the introduction of agile methodology. In order to make certain that these publications were captured, we also included synonyms to ‘agile transition’ to the search string. The final search string looked like this: \emph{(("organizational change" OR "organisational change" OR "organizational development" OR "organisational development") OR ((agile OR kanban) AND (transition OR adaptation OR employment))) AND ("software engineering" OR "software development" OR "agile development" OR "software engineer" OR "software developer" OR "software project")}

The quality of the search strings was verified by a pilot search for three known organizational change related publication~\cite{cohn2003introducing, smits2011agile, mockus2010organizational}. The search string caught all of them.

\subsection{Research Selection Criteria}
\label{sec_selection_criteria}
In order to reduce the likelihood of bias, the study selection criteria were derived. The criteria were intended to identify those primary publications that provide direct evidence regarding the aim of the systematic literature review~\cite{kitchenham2004procedures}.

\begin{enumerate}

\item[] Inclusion Criteria
\begin{enumerate}
\item[] \emph{Publication Year:} We limited the search to include publications between January 2000 and July 2015 The start date was set in order to capture studies related to the agile software development approach, which has had a major influence on Software Engineering.
\item[] \emph{Publication Type:} We choose to include papers published both in journals and in conference proceedings.
\item[] \emph{Content:} We included publication if they were related to both software engineering and organizational change, i.e. the changes had to have been studied in relation software engineering activities or to software engineers.
\end{enumerate}

\item[] Exclusion Criteria
\begin{enumerate}
\item[] \emph{Language:} We limited this study to only include papers written in English.
\item[] \emph{Publication Type:} We excluded papers where we could not locate a full paper version, although only one paper was affected by this exclusion criterion.
\end{enumerate}

\end{enumerate}

\subsection{Research Selection Process}
\label{sec_research_selection_process}

\begin{figure}[ht]
\centering
\includegraphics[width=1\linewidth]{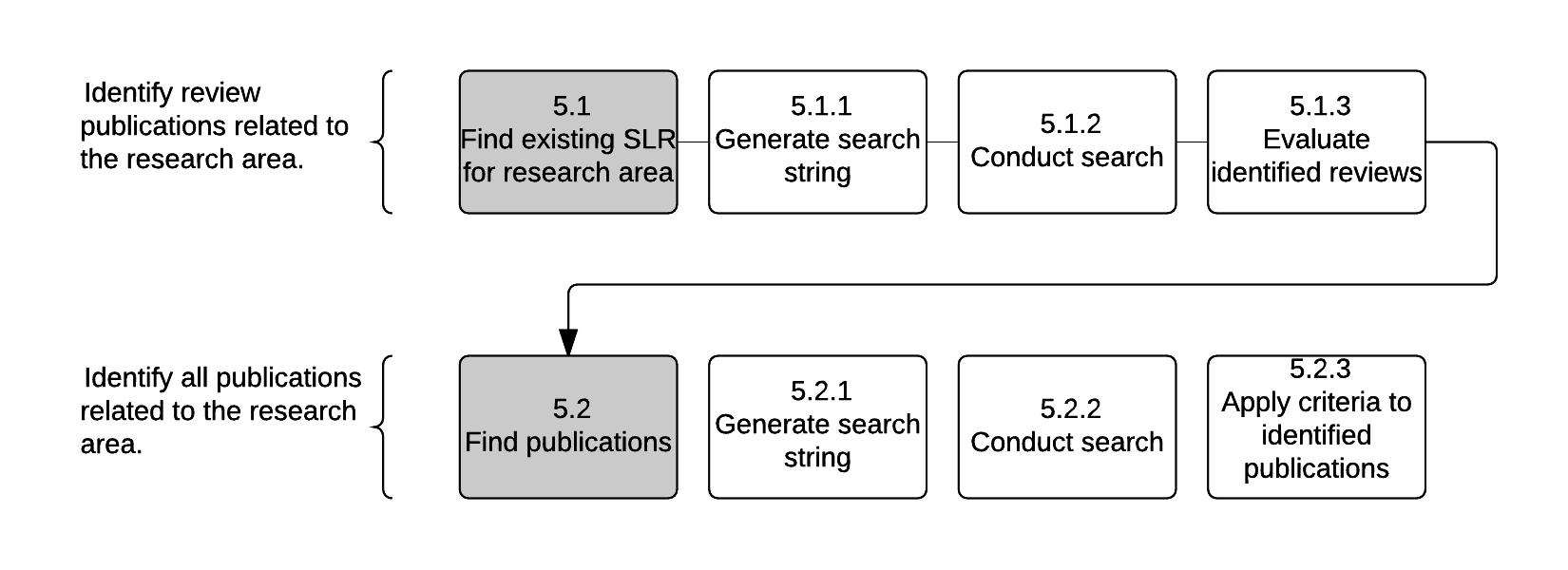}
\caption{Steps included in the selection process.}
\label{fig:method_selection}
\end{figure}

As shown in Figure~\ref{fig:method_selection} the research selection included two main steps. In the first step we identified if there existed any reviews that covered a major part of organizational change in software engineering. Since no such reviews were found, we identified none-review studies. These main steps are detailed below.

\begin{enumerate}
\item[5.1] First, we explore if any systematic literature reviews existed covering a substantial part of organizational change in software engineering.
\begin{enumerate}
\item[5.1.1] The search string was generated by combining the search string defined in section~\ref{sec_search_string} above with synonyms for ‘systematic review’ (defined by Biolchini et al.~\cite{biolchini2005systematic}) with a logical AND operator.
\item[5.1.2] Next, the search string was applied to the selected data sources.
\item[5.1.3] The search identified eleven literature reviews, however, four of these were rejected based on review of title and abstract (see table~\ref{table:No_Search_Results}). As mentioned in section~\ref{sec_methods_the_need}, the included reviews considered and analyzed different aspects and concepts of organizational change in software engineering, e.g. agile development~\cite{mohanarajah2015improved, razavi2014agile,eck2014fit}, software process improvement\cite{unterkalmsteiner2012evaluation,lavallee2012impacts,pino2008software}. Hence, the search did not identify any review covering organizational change in software engineering as a whole.
\end{enumerate}

\item[5.2] Next, we identified organizational change publications related to software engineering.
\begin{enumerate}
\item[5.2.1] The research sting used are show in section~\ref{sec_search_string}.
\item[5.2.2] Next, the search string was applied to the selected data sources.
\item[5.2.3] The criteria used in the evaluation are defined in section~\ref{sec_selection_criteria}. As suggested by Kitchenham~\cite{kitchenham2004procedures} the selection criteria were initially interpreted liberally, so that unless the identified systematic reviews could be clearly excluded based on titles and abstracts, full copies were obtained. In total, 41 primary publications were identified and included in this review (see table~\ref{table:No_Search_Results}).
\end{enumerate}
\end{enumerate}

\begin{table}[ht]
\footnotesize
\centering
\begin{tabular}{| C{.15\textwidth} | C{.25\textwidth} | C{.12\textwidth} | C{.12\textwidth} |} \hline
\rowcolor{black!15}\Centering Search String & \Centering Before review & \Centering Abstract and title review & \Centering Full review \\ \hline
5.1 &   11 (1,0,8,2)             & 7     & 7 \\ \hline
5.2 &   778 (15, 196, 381, 186)  & 85    & 41 \\ \hline
\end{tabular}
\caption{Number of search results at the three different review stages. The numbers within brackets in column 'Before review' show the search results per database - PsycInfo, IEEE, Scopus and Web of science.}
\label{table:No_Search_Results}
\end{table}

\subsection{Data Extraction and Synthesis}
The properties to extract for each included publications were chosen to meet the primary purpose of the literature review. Table~\ref{table:properties} shows the complete list of properties (P1-P4) each with a brief description, and, where applicable, its possible value range.

To get an overview of the research area and identify temporal trends we extracted the publication year (P1). Three properties, i.e. P2-P3, extracted information regarding \emph{what} had been studied, i.e.  what human factors or concepts, if any, were analyzed or considered in the study (P2) and what type of organizational change was analyzed in the study (P3).

One property (P4) extracted information related to \emph{how}, or in what way, the studies had been conducted, i.e. the faculty affiliation of the researcher(s).

In order to verify the reliability of the extraction process a pilot extraction was conducted, where three researchers classified six randomly selected publications~\cite{brereton2007lessons}. Comparing the results from the researchers showed that the extraction process was straightforward for all properties except.

\end{document}